\documentclass{article}

\usepackage{arxiv}
\usepackage[square,numbers]{natbib}
\usepackage{graphicx}
\usepackage{amsmath}
\usepackage{amssymb}
\usepackage{setspace}
\usepackage{textcomp}

\usepackage[utf8]{inputenc} 
\usepackage[T1]{fontenc}    
\usepackage{hyperref}       
\usepackage{url}            
\usepackage{booktabs}       
\usepackage{amsfonts}       
\usepackage{nicefrac}       
\usepackage{microtype}      
\usepackage{lipsum}
\usepackage{multirow}
\usepackage{graphicx}
\usepackage{array}
\graphicspath{ {./images/} }

\title{Block-diagonalizable two-dimensional generalized Ising systems (BD2DGIS): the eigenvalues and eigenvectors}

\author{Vadym Sakhno \And Mykola Sakhno}

\begin{document}

\maketitle

\begin{abstract}

This paper is a continuation of \cite{sakhno2020matrix} and \cite{sakhno2021free}, where the block-diagonalizable two-dimensional generalized Ising systems (BD2DGIS) were introduced. In this paper, their eigenvalues, eigenvectors and Jordan normal form are analyzed in detail using the simplest quantum field model. 

\end{abstract}

\section{Introduction} \label{sec:Introduction}

Since this paper is a continuation of \cite{sakhno2020matrix} and \cite{sakhno2021free}, the Introduction provides detailed summaries of \cite{sakhno2020matrix} and \cite{sakhno2021free}.

\subsection{Summary of [1]}

\begin{figure}[ht!] 
    \centering
    \includegraphics[width=\textwidth]{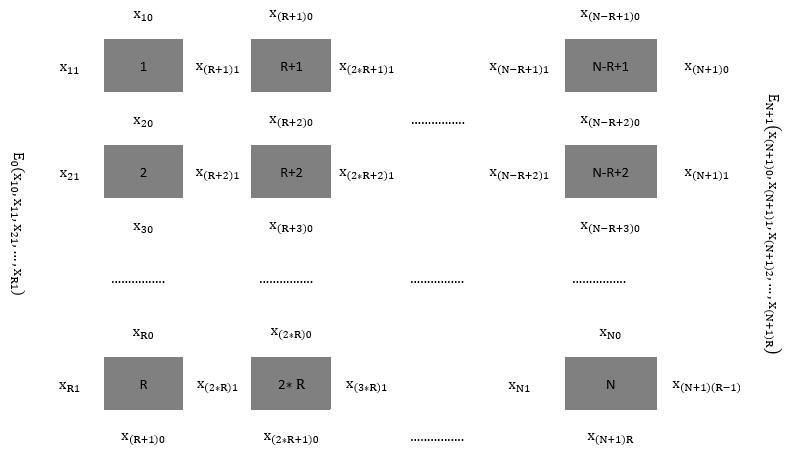}
    \caption{The 2D system under consideration.}
    \label{fig:system}
\end{figure}

In \cite{sakhno2020matrix}, the Ising model was generalized to a system of cells interacting exclusively by presence of shared spins. Within the cells there were interactions of any complexity, the simplest intracell interactions came down to the Ising model. The approach was developed to constructing the exact matrix model for any considered system in the simplest way. Using the approach, the exact matrix model for a two-dimensional generalized Ising model was constructed. The 2D system under consideration is shown in Figure \ref{fig:system} (see Figure 1 of \cite{sakhno2020matrix}). 

The 2D system of Figure \ref{fig:system} consists of \( (N+2) \) cells having spins with two values \( \{-\frac{1}{2},\frac{1}{2}\} \): $N$ internal cells numbered from $1$ to $N$, start cell numbered $0$, and finish cell numbered $(N+1)$. $N$ internal cells form $R$ rows. The lowest spin of a column continues into the highest spin of the next column (see \( x_{\left(R+1\right)0} \) in Figure \ref{fig:system}). Thus, cells are placed along a helix. Let the first column be completed, and the last column may be uncompleted. Each internal cell $n$ has four spins, its energy is a given function $E_n\left(x_{n0},\ x_{n1},x_{(n+1)0},x_{(n+R)1}\right)$. Substituting it into (3) of \cite{sakhno2020matrix}, the internal cell function $Z_n\left(x_{n0},\ x_{n1},x_{(n+1)0},x_{(n+R)1}\right)$ is

\begin{equation} \label{eq:Z_n}
        Z_n\left(x_{n0},\ x_{n1},x_{(n+1)0},x_{(n+R)1}\right) = \exp{ \left( -\frac{E_n\left(x_{n0},\ x_{n1},x_{(n+1)0},x_{(n+R)1}\right)}{k_B T} \right) } > 0.
\end{equation}

In \cite{sakhno2020matrix}, for each spin \( x_{n\nu} \in \{-\frac{1}{2}, \ +\frac{1}{2} \} \), the substituting spin-number \( i_{n\nu} \in \{ 0, 1 \} \) was introduced according to (6) of \cite{sakhno2020matrix}

\begin{equation} \label{eq:x_n_nu}
    x_{n\nu}=i_{n\nu}-\frac{1}{2}.
\end{equation}

In the internal cell function (\ref{eq:Z_n}), the substitution of each spin with its spin-number yielded the internal cell frame, which was a set of 16 values, numbered with a compound number of spin-numbers. 
The internal cells in \cite{sakhno2020matrix} may vary, but in this paper they are similar. Therefore, it suffices to consider the first cell $n=1$. Its frame is (17) of \cite{sakhno2020matrix}

\begin{equation} \label{eq:Z_1}
    Z_{1i_{10}i_{11}i_{20}i_{\left(R+1\right)1}}=Z_1\left(i_{10}-\frac{1}{2}, \ i_{11}-\frac{1}{2}, \ i_{20}-\frac{1}{2}, i_{\left(R+1\right)1}-\frac{1}{2}\right).
\end{equation}

 In \cite{sakhno2020matrix}, four $2^R\times2^R$ block-diagonal matrices were constructed from the frame, along the diagonal of which there were identical $2\times2$ blocks. Their elements with compound row number  $i_1 i_2 \ldots i_R$ were non-zero only if all row sub-numbers except the last sub-number $R$ were equal to the corresponding column sub-numbers. To emphasize this, these $2^R\times2^R$ block-diagonal matrices were denoted as $2\times2$ matrices with the number $[R]$ (see (25) of \cite{sakhno2020matrix}), for example 

\begin{equation} \label{eq:mat_Z_11_R}
    \begin{aligned}
    \begin{pmatrix}
        Z_{10000} & Z_{10100} \\
        Z_{10001} & Z_{10101}
    \end{pmatrix}
    _{[R]}
    =
    \begin{pmatrix}
        Z_{10000} & Z_{10100} & 0 & 0 & \ldots & 0 & 0 \\
        Z_{10001} & Z_{10101} & 0 & 0 & \ldots & 0 & 0 \\
        0 & 0 & Z_{10000} & Z_{10100} & \ldots & 0 & 0 \\
        0 & 0 & Z_{10001} & Z_{10101} & \ldots & 0 & 0 \\
        \ldots & \ldots & \ldots & \ldots & \ldots & \ldots & \ldots \\
        0 & 0 & 0 & 0 & \ldots & Z_{10000} & Z_{10100} \\
        0 & 0 & 0 & 0 & \ldots & Z_{10001} & Z_{10101}
    \end{pmatrix}
    .
    \end{aligned}
\end{equation}

Also $2^R\times2^R$ cyclic shift matrix was introduced in (23) of \cite{sakhno2020matrix}

\begin{equation} \label{eq:mat_P_r}
\setcounter{MaxMatrixCols}{12}
    \begin{aligned}
        \overleftrightarrow{P_r}
        & =
        \begin{pmatrix}
            1 & 0 & \ldots & 0 & 0 & \ldots & 0 & 0 & \ldots & 0 & 0 & \ldots \\
            0 & 0 & \ldots & 0 & 0 & \ldots & 1 & 0 & \ldots & 0 & 0 & \ldots \\
            0 & 1 & \ldots & 0 & 0 & \ldots & 0 & 0 & \ldots & 0 & 0 & \ldots \\
            0 & 0 & \ldots & 0 & 0 & \ldots & 0 & 1 & \ldots & 0 & 0 & \ldots \\
            \ldots & \ldots & \ldots & \ldots & \ldots & \ldots & \ldots & \ldots & \ldots & \ldots & \ldots & \ldots \\
            \ldots & \ldots & \ldots & \ldots & \ldots & \ldots & \ldots & \ldots & \ldots & \ldots & \ldots & \ldots \\
            0 & 0 & \ldots & 1 & 0 & \ldots & 0 & 0 & \ldots & 0 & 0 & \ldots \\
            0 & 0 & \ldots & 0 & 0 & \ldots & 0 & 0 & \ldots & 1 & 0 & \ldots \\
            0 & 0 & \ldots & 0 & 1 & \ldots & 0 & 0 & \ldots & 0 & 0 & \ldots \\
            0 & 0 & \ldots & 0 & 0 & \ldots & 0 & 0 & \ldots & 0 & 1 & \ldots \\
            \ldots & \ldots & \ldots & \ldots & \ldots & \ldots & \ldots & \ldots & \ldots & \ldots & \ldots & \ldots \\
            \ldots & \ldots & \ldots & \ldots & \ldots & \ldots & \ldots & \ldots & \ldots & \ldots & \ldots & \ldots
        \end{pmatrix}
        .
    \end{aligned}
\end{equation}

Then the sought-for $2^{R+1}\times2^{R+1}$ internal cell matrix was represented in (24) of \cite{sakhno2020matrix} as

\begin{equation} \label{eq:mat_Z_1}
    \overleftrightarrow{Z_1}
    =
    \begin{pmatrix}    
       \begin{pmatrix}
           Z_{10000} & Z_{10100} \\
           Z_{10001} & Z_{10101}
       \end{pmatrix}
       _{[R]}
    &
       \begin{pmatrix}
           Z_{11000} & Z_{11100} \\
           Z_{11001} & Z_{11101}
       \end{pmatrix}
       _{[R]}    
    \\
       \begin{pmatrix}
           Z_{10010} & Z_{10110} \\
           Z_{10011} & Z_{10111}
       \end{pmatrix}
       _{[R]}
    &
       \begin{pmatrix}
           Z_{11010} & Z_{11110} \\
           Z_{11011} & Z_{11111}
       \end{pmatrix}
       _{[R]}    
    \end{pmatrix}
    *
    \begin{pmatrix}
        \overleftrightarrow{P_r} & 0 \\
        0 & \overleftrightarrow{P_r}
    \end{pmatrix}
    ,
\end{equation}

As shown in Figure \ref{fig:system}, the start cell energy is $E_0 \left( x_{10}, \ x_{11}, \ x_{21}, \ldots, x_{R1}  \right)$ and the finish cell energy is $E_{N+1} \left(  x_{(N+1)0}, x_{(N+1)1}, x_{(N+1)2}, \ldots, x_{(N+1)R}  \right)$, which can be any given functions of $(R + 1)$ spins. Substituting them into (3) of \cite{sakhno2020matrix} gives the start cell function and the finish cell function (see (\ref{eq:Z_n}) for an internal cell). Then substituting each spin with its spin-number according to (\ref{eq:x_n_nu}) gives the start cell frame $Z_{0i_{10}i_{11}i_{21} \ldots i_{R1}}$ and the finish cell frame $Z_{(N+1)i_{(N+1)0} \ldots i_{(N+1)(R-1)}i_{(N+1)R}}$ of $2^{R+1}$ values each. The start cell vector $\overleftarrow{Z_0}$ is $2^{R+1}$ column vector of the start cell frame values and the finish cell vector $\overrightarrow{Z_{N+1}}$ is $2^{R+1}$ row vector of the finish cell frame values, let them be called boundary conditions

\begin{equation} \label{eq:vec_Z_0_N}
   \overleftarrow{Z_0} 
   = \overleftarrow{\|Z_{0i_{10}i_{11}i_{21} \ldots i_{R1}}\|} \ ,
   \quad
   \overrightarrow{Z_{N+1}} 
   = \overrightarrow{\|Z_{(N+1)i_{(N+1)0} \ldots i_{(N+1)(R-1)}i_{(N+1)R}}\|} \ .
\end{equation}

Then the resulting exact partition function $Z$ for $N$ identical internal cell matrices $\overleftrightarrow{Z_1}$ was (9) in \cite{sakhno2020matrix}

\begin{equation} \label{eq:Z}
    Z=\overrightarrow{Z_{N+1}} * {\overleftrightarrow{Z_1}}^N * \overleftarrow{Z_0}.
\end{equation}

\subsection{Summary of [2]}

\cite{sakhno2021free} is a continuation of \cite{sakhno2020matrix}.

The properties of internal cell matrix (\ref{eq:mat_Z_1}) for light block-diagonalization were specified in (9) of \cite{sakhno2021free}

\begin{equation} \label{eq:props_for_diag}
\begin{aligned}
   \begin{pmatrix}
      Z_{10000} & Z_{11000} \\
      Z_{10010} & Z_{11010}
   \end{pmatrix}, \quad
   \begin{pmatrix}
      Z_{10100} & Z_{11100} \\
      Z_{10110} & Z_{11110}
   \end{pmatrix}, \quad
   \begin{pmatrix}
      Z_{10001} & Z_{11001} \\
      Z_{10011} & Z_{11011}
   \end{pmatrix}, \quad
   \begin{pmatrix}
      Z_{10101} & Z_{11101} \\
      Z_{10111} & Z_{11111}
   \end{pmatrix}\\
are \quad diagonalized \quad by \quad the \quad same \quad similarity \quad transformation.
\end{aligned}
\end{equation}.

And an example of BD2DGIS was given in Table 1 and Figure 2 of \cite{sakhno2021free}.

The exact internal cell matrix (\ref{eq:mat_Z_1}) for the example was obtained in (11) of \cite{sakhno2021free}

\begin{equation} \label{eq:mat_Z_1_ex}
    \overleftrightarrow{Z_1}
    = \zeta *
    \begin{pmatrix}    
       \begin{pmatrix}
           13 & 69 \\
           68 & 64
       \end{pmatrix}
       _{[R]}
    &
       \begin{pmatrix}
           4 & 22 \\
           24 & 22
       \end{pmatrix}
       _{[R]}    
    \\
       \begin{pmatrix}
           4 & 22 \\
           24 & 22
       \end{pmatrix}
       _{[R]}
    &
       \begin{pmatrix}
           7 & 36 \\
           32 & 31
       \end{pmatrix}
       _{[R]}    
    \end{pmatrix}
    *
    \begin{pmatrix}
        \overleftrightarrow{P_r} & 0 \\
        0 & \overleftrightarrow{P_r}
    \end{pmatrix},
\end{equation}

where

\begin{equation} \label{eq:zeta}
   \zeta = \cfrac{1}{2*\sqrt[4]{11}*\sqrt[16]{2^{13}*3^5*1103011}} \approx 0.0465.
\end{equation}

Then $2^{R+1}\times2^{R+1}$ matrix $\overleftrightarrow{S_0}$ was introduced in (19) of \cite{sakhno2021free}

\begin{equation} \label{eq:mat_S_0}
    \overleftrightarrow{S_0}
    =
    \begin{pmatrix}
      2*\overleftrightarrow{1} & -\overleftrightarrow{1} \\
      \overleftrightarrow{1} & 2*\overleftrightarrow{1}
    \end{pmatrix},
\end{equation}

where $\overleftrightarrow{1}$ is $2^R\times2^R$ identity matrix.

And a similarity transformation with matrix $\overleftrightarrow{S_0}$ of (\ref{eq:mat_S_0}) was performed over the internal cell matrix $\overleftrightarrow{Z_1}$ of (\ref{eq:mat_Z_1_ex}) in (20) of \cite{sakhno2021free}

\begin{equation} \label{eq:mat_Z_1_S_0}
    \overleftrightarrow{Z_{1S_0}}
    = {\overleftrightarrow{S_0}}^{-1} * \overleftrightarrow{Z_1} * \overleftrightarrow{S_0} =
    \begin{pmatrix}    
      \overleftrightarrow{B_0}_{[R]} & \overleftrightarrow{0}
    \\
      \overleftrightarrow{0} & \overleftrightarrow{B_1}_{[R]}
    \end{pmatrix}
    *
    \begin{pmatrix}
        \overleftrightarrow{P_r} & 0 \\
        0 & \overleftrightarrow{P_r}
    \end{pmatrix},
\end{equation}

where

\begin{equation} \label{eq:mat_B_0_B_1_R}
   \overleftrightarrow{B_0}_{[R]} = \zeta *
      \begin{pmatrix}
          15 & 80 \\
          80 & 75
      \end{pmatrix}_{[R]}, \quad
   \overleftrightarrow{B_1}_{[R]} = \zeta *
      \begin{pmatrix}
          5 & 25 \\
          20 & 20
      \end{pmatrix}_{[R]}.
\end{equation}

The resulting matrix $\overleftrightarrow{Z_{1S_0}}$ is the product of two matrices, each of which is block-diagonal of two $2^R\times2^R$ blocks.

Then $2^R\times2^R$ matrices were introduced for any $\rho \in[1,R]$ in (30) of \cite{sakhno2021free}

\begin{equation} \label{eq:mat_S_1_S_2_rho}
    \overleftrightarrow{S_1}_{[\rho]} =
    \begin{pmatrix}
      -3+\sqrt{73} & -3-\sqrt{73} \\
      8 & 8
  \end{pmatrix}
  _{[\rho]}, \quad
    \overleftrightarrow{S_2}_{[\rho]} =
    \begin{pmatrix}
      -3+\sqrt{89} & -3-\sqrt{89} \\
      8 & 8
  \end{pmatrix}
  _{[\rho]}.
\end{equation}

Similarity transformation with matrices $\overleftrightarrow{S_1}_{[\rho]}$ and $\overleftrightarrow{S_2}_{[\rho]}$ of (\ref{eq:mat_S_1_S_2_rho}) over matrices $\overleftrightarrow{B_0}_{[\rho]}$ and $\overleftrightarrow{B_1}_{[\rho]}$ of (\ref{eq:mat_B_0_B_1_R}) diagonalized the latter in (31) of \cite{sakhno2021free}

\begin{equation} \label{eq:mat_B_0_B_1_R_diag}
\begin{aligned}
   \overleftrightarrow{B_0}_{[\rho]} = &\overleftrightarrow{S_1}_{[\rho]}
      * \left( {\overleftrightarrow{S_1}_{[\rho]}}^{-1} * \overleftrightarrow{B_0}_{[\rho]} 
      * \overleftrightarrow{S_1}_{[\rho]} \right) * {\overleftrightarrow{S_1}_{[\rho]}}^{-1}
      = \lambda_1 * \overleftrightarrow{S_1}_{[\rho]} *
      \begin{pmatrix}
        1 & 0 \\
        0 & \lambda_{21}
      \end{pmatrix}_{[\rho]}
      * {\overleftrightarrow{S_1}_{[\rho]}}^{-1},
     \\
   \overleftrightarrow{B_1}_{[\rho]} = &\overleftrightarrow{S_2}_{[\rho]}
      * \left( {\overleftrightarrow{S_2}_{[\rho]}}^{-1} * \overleftrightarrow{B_1}_{[\rho]} 
      * \overleftrightarrow{S_2}_{[\rho]} \right) * {\overleftrightarrow{S_2}_{[\rho]}}^{-1}
      = \lambda_3 * \overleftrightarrow{S_2}_{[\rho]} *
      \begin{pmatrix}
        1 & 0 \\
        0 & \lambda_{43}
      \end{pmatrix}_{[\rho]}
      * {\overleftrightarrow{S_2}_{[\rho]}}^{-1},
\end{aligned} 
\end{equation}

where taking into account (\ref{eq:zeta})

\begin{equation} \label{eq:lambda}
\begin{aligned}
   \lambda_1 = \ &\zeta * \left( 45+10\sqrt{73} \right) \approx 6.063, \quad
   \lambda_{21} = \cfrac{9-2\sqrt{73}}{9+2\sqrt{73}} \approx -0.3100,
   \\
   \lambda_3 = \ &\zeta * \cfrac{25+5\sqrt{89}}{2} \approx 1.677, \quad
   \lambda_{43} = \cfrac{5-\sqrt{89}}{5+\sqrt{89}} \approx -0.3072.
\end{aligned}
\end{equation}

\section{Constructing the model} \label{sec:model}

\subsection{Quasi-diagonalization of the internal cell matrix}

Let $2^R\times2^R$ matrices $\overleftrightarrow{S_3}$ and $\overleftrightarrow{S_4}$ be constructed of matrices (\ref{eq:mat_S_1_S_2_rho}) as follows

\begin{equation} \label{eq:mat_S_3_S_4}
    \overleftrightarrow{S_3} = \prod_{\rho=1}^{R} \overleftrightarrow{S_1}_{[\rho]}, \quad
    \overleftrightarrow{S_4} = \prod_{\rho=1}^{R} \overleftrightarrow{S_2}_{[\rho]}.
\end{equation}

Matrices $\overleftrightarrow{S_3}$ and $\overleftrightarrow{S_4}$ commute with matrix $P_r$ of (\ref{eq:mat_P_r}). And similarity transformation with them diagonalizes respectively matrices $\overleftrightarrow{B_0}_{[R]}$ and $\overleftrightarrow{B_1}_{[R]}$ of (\ref{eq:mat_B_0_B_1_R}). 

Let $2^{R+1}\times2^{R+1}$ matrix $\overleftrightarrow{S_5}$ be constructed of matrices (\ref{eq:mat_S_3_S_4}) as follows

\begin{equation} \label{eq:mat_S_5}
    \overleftrightarrow{S_5} = 
    \begin{pmatrix}    
      \overleftrightarrow{S_3} & \overleftrightarrow{0}
    \\
      \overleftrightarrow{0} & \overleftrightarrow{S_4}
    \end{pmatrix}.
\end{equation}

The similarity transformation with matrix $\overleftrightarrow{S_5}$ of (\ref{eq:mat_S_5}) over the block-diagonalized internal cell matrix $\overleftrightarrow{Z_{1S_0}}$ of (\ref{eq:mat_Z_1_S_0}) taking into account (\ref{eq:mat_B_0_B_1_R_diag}) gives the quasi-diagonal matrix

\begin{equation} \label{eq:mat_Z_1_S_0_S_5}
    \overleftrightarrow{Z_{1S_0S_5}} = {\overleftrightarrow{S_5}}^{-1} * \overleftrightarrow{Z_{1S_0}} * \overleftrightarrow{S_5} = 
    \begin{pmatrix}    
      \lambda_1 * {\overleftrightarrow{\lambda}}_{[R]} \left( \lambda_{21} \right) & \overleftrightarrow{0}
    \\
      \overleftrightarrow{0} & \lambda_3 * {\overleftrightarrow{\lambda}}_{[R]} \left( \lambda_{43} \right)
    \end{pmatrix}
    *
    \begin{pmatrix}
        \overleftrightarrow{P_r} & 0 \\
        0 & \overleftrightarrow{P_r}
    \end{pmatrix}.
\end{equation}

where

\begin{equation} \label{eq:mat_Lam_R}
    {\overleftrightarrow{\lambda}}_{[R]} \left( \lambda \right) = 
      \begin{pmatrix}
        1 & 0 \\
        0 & \lambda
      \end{pmatrix}_{[R]},
      \quad
      |\lambda| < 1.
\end{equation}

\subsection{The simplest quantum field model} \label{subsec:QuantumField}

The initial task is to diagonalize the transformed internal cell $2^{R+1}\times2^{R+1}$ matrix $\overleftrightarrow{Z_{1S_0S_5}}$ of (\ref{eq:mat_Z_1_S_0_S_5}), which boils down to diagonalization of the following $2^R\times2^R$ matrix

\begin{equation} \label{eq:mat_Z}
    \overleftrightarrow{Z} \left( \lambda \right) = {\overleftrightarrow{\lambda}}_{[R]} \left( \lambda \right) * \overleftrightarrow{P_r}.
\end{equation}

Let $2^R\times2^R$ matrix $\overleftrightarrow{Z}$ of (\ref{eq:mat_Z}) acts on $2^R$ column vector $\overleftarrow{V}$, similar to $\overleftarrow{Z_0}$ of (\ref{eq:vec_Z_0_N})

\begin{equation} \label{eq:vec_V}
  \overleftarrow{V} 
  = \overleftarrow{\|v_{\overline{\theta}}\|}
  = \overleftarrow{\|v_{\theta_1\theta_2 \ldots \theta_R}\|},
  \quad
  \theta_1 \in [0, 1], \ \theta_2 \in [0, 1], \ \ldots, \ \theta_R \in [0, 1],
\end{equation}

where $\overline{\theta} = \theta_1\theta_2 \ldots \theta_R \in [0, 2^R-1]$ - the compound number of a vector element, consisting of subnumbers $\theta_1,\theta_2, \ldots \theta_R$.

Let $2^R$ basis $W_R$ of $2^R$ basis column vectors $\overleftarrow{W_{q_1q_2 \ldots q_R}}=\overleftarrow{W_{\overline{q}}}$ be introduced, in which each basis vector is numbered with a compound number $\overline{q} \in [0, 2^R-1]$, consisting of subnumbers $q_1 \in [0,1], \ q_2 \in [0,1], \ \ldots, \ q_R \in [0,1]$

\begin{equation} \label{eq:vec_W}
  \overleftarrow{W_{\overline{q}}}
  = \overleftarrow{\|w_{\overline{q}, \overline{\theta}}\|},
  \quad
  \overline{\theta} \in [0, 2^R-1]\text{ - the compound number of a vector element},
\end{equation}

where

\begin{equation*}
    w_{\overline{q}, \overline{\theta}} =
    \begin{cases}
    1& \text{if $\overline{\theta} = \overline{q}$}, \\
    0& \text{if $\overline{\theta} \neq \overline{q}$}.
    \end{cases}
\end{equation*}

For example, for $R = 2$ the basis $W_2$ consists of $2^2=4$ basis column 4-vectors

\begin{equation*}
   W_{00} = 
   \begin{pmatrix}
     1 \\ 0 \\ 0 \\ 0 
   \end{pmatrix},
   W_{01} = 
   \begin{pmatrix}
     0 \\ 1 \\ 0 \\ 0 
   \end{pmatrix},
   W_{10} = 
   \begin{pmatrix}
     0 \\ 0 \\ 1 \\ 0 
   \end{pmatrix},
   W_{11} = 
   \begin{pmatrix}
     0 \\ 0 \\ 0 \\ 1 
   \end{pmatrix}.
\end{equation*}

Then the vector $\overleftarrow{V}$ of (\ref{eq:vec_V}) may be represented by coordinates $v$ with respect to the basis $W_R$

\begin{equation} \label{eq:vect_V_by_W}
  \overleftarrow{V} 
  = \sum_{q_1,q_2, \ldots, q_R \in [0, 1]} 
  v_{q_1q_2 \ldots q_R} *
  \overleftarrow{W_{q_1q_2 \ldots q_R}}
\end{equation}

For equations (\ref{eq:mat_Z}) $\ldots$ (\ref{eq:vect_V_by_W}), a quantum field model may be constructed that greatly simplifies realization. 

\begin{figure}[ht!] 
    \centering
    \includegraphics[width=0.5\textwidth]{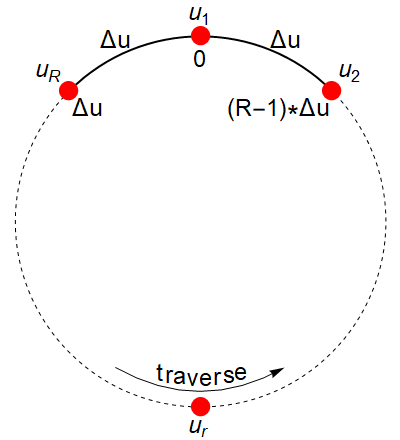}
    \caption{The circle-space $u$.}
    \label{fig:circle}
\end{figure}

Let the space $u$ be introduced being a circle with $R$ points $u_1, \ u_2, \ u_3, \ \ldots, \ u_r, \ \ldots, \ u_{R-1}, \ u_R$. Let the distance between the nearest points $\Delta u$ of the circle-space $u$ be called the space step. Let the coordinate of point $u_1$ be $0$, then coordinates of points $u_R, \ u_{R-1}, \ \ldots, \ u_2, \ u_1$ at counterclockwise traversing are $ \Delta u, \ 2 * \Delta u, \ \ldots, \ \left( R - 2 \right) * \Delta u, \ \left( R - 1 \right) * \Delta u$ respectively  (see Figure \ref{fig:circle}).

Let two basis quantum states be possible at each point $u_{r}$: a quantum is absent or a quantum is present, which is described by a two-valued quantum variable $q_{r} \in [0, 1]$. Then $2^R$ basis quantum states for the quantum field of the entire circle-space $\overleftarrow{W_{q_1q_2 \ldots q_R}}$ may be introduced according to (\ref{eq:vec_W}). And any state of the quantum field $\overleftarrow{V}$ (see (\ref{eq:vec_V})) is their superposition (see (\ref{eq:vect_V_by_W})), let the set of these states be called the state-space.

The state $\overleftarrow{V}$ is acted $N$ times (see (\ref{eq:Z})) by the operator $\overleftrightarrow{Z}$ of (\ref{eq:mat_Z}). Let one action of operator $\overleftrightarrow{Z}$ be called an elementary action. Let a change in the state of the quantum field due to the elementary action be called motion per time step $\Delta t$.

The task of this paper is to find and analyze eigenvalues and eigenvectors of the elementary action $\overleftrightarrow{Z}$ of (\ref{eq:mat_Z}).

\section{The eigenvalues of the elementary action} \label{sec:eigenvalues}

\subsection{The general analysis} \label{subsec:eigenval_gen}

The elementary action $\overleftrightarrow{Z}$ of (\ref{eq:mat_Z}) is the product of two matrices. The cyclic shift matrix $\overleftrightarrow{P_r}$ acts on any basis vector $\overleftarrow{W_{q_1q_2 \ldots q_R}}$ of (\ref{eq:vec_W}), shifting its sub-numbers cyclically as follows

\begin{equation} \label{eq:mat_P_by_W}
  \overleftrightarrow{P_r} *
  \overleftarrow{W_{q_1 q_2 \ldots q_R}}
  = \overleftarrow{W_{q_2 \ldots q_R q_1}}.
\end{equation}
 
And the diagonal matrix ${\overleftrightarrow{\lambda}}_{[R]} \left( \lambda \right)$ of (\ref{eq:mat_Lam_R}) acts on any basis vector $\overleftarrow{W_{q_1q_2 \ldots q_R}}$ of (\ref{eq:vec_W}), leaving it unchanged and only multiplying by the coefficient

\begin{equation} \label{eq:mat_Lam_by_W}
  {\overleftrightarrow{\lambda}}_{[R]} \left( \lambda \right) *
  \overleftarrow{W_{q_1q_2 \ldots q_R}}
  = 
  \begin{cases}
    \overleftarrow{W_{q_1q_2 \ldots q_R}}& \text{if $q_R = 0$}, \\
    \lambda * \overleftarrow{W_{q_1q_2 \ldots q_R}}& \text{if $q_R = 1$}.
    \end{cases}
\end{equation}

Hence, for any $R$, there are two singleton invariant subspaces of the elementary action $\overleftrightarrow{Z}$ of (\ref{eq:mat_Z}):

\begin{itemize}
    \item Subspace called the vacuum with the vacuum eigenvector $\overleftarrow{W_{00 \ldots 0}}$ and eigenvalue $\Lambda \left( 00 \ldots 0 \right) = 1$
    
\begin{equation} \label{eq:vacc}
  \overleftrightarrow{Z} * \overleftarrow{W_{00 \ldots 0}}
  = \Lambda \left( 00 \ldots 0 \right) * \overleftarrow{W_{00 \ldots 0}} 
  = 1 * \overleftarrow{W_{00 \ldots 0}}. 
\end{equation}

    It will be shown later in this subsection that this eigenvalue is greater in absolute value than any other eigenvalue.

    \item Subspace called the full with the full eigenvector $\overleftarrow{W_{11 \ldots 1}}$ and eigenvalue $\Lambda \left( 11 \ldots 1 \right) = \lambda, \quad |\lambda| < 1$
    
\begin{equation} \label{eq:filled}
  \overleftrightarrow{Z} * \overleftarrow{W_{11 \ldots 1}}
  = \Lambda \left( 11 \ldots 1 \right) * \overleftarrow{W_{11 \ldots 1}}
  = \lambda * \overleftarrow{W_{11 \ldots 1}}.   
\end{equation}

    It will be shown later in this subsection that this eigenvalue is smaller in absolute value than any other eigenvalue.

\end{itemize}

Each of the other basis vectors $\overleftarrow{W_{q_1q_2 \ldots q_R}}$ is an eigenvector of the elementary action $\overleftrightarrow{Z}$ powered $R$

\begin{equation} \label{eq:mat_Z_by_R}
  \overleftrightarrow{Z}^R *
  \overleftarrow{W_{q_1q_2 \ldots q_R}}
  = \lambda^Q * \overleftarrow{W_{q_1q_2 \ldots q_R}},
\end{equation}

where $Q$ is the amount of quanta in the basis vector $\overleftarrow{W_{q_1q_2 \ldots q_R}}$

\begin{equation} \label{eq:Q}
    Q = q_1 + q_2 + \ldots + q_R, \quad 0 \leq Q \leq R.
\end{equation}

Now one can divide the state-space into subspaces according to the values of $Q$. Each subspace has the amount of basis vectors equal to the binomial coefficients $\displaystyle \binom{R}{Q} = \frac{R!}{Q!*(R-Q)!}$. And eigenvalues $\Lambda \left( q_1 q_2 \ldots q_R \right) = \Lambda \left( \overline{q} \right)$ satisfying, based on (\ref{eq:mat_Z_by_R}), the following equation

\begin{equation} \label{eq:Lam_in_R}
  \Lambda^R \left( \overline{q} \right) = \lambda^Q.
\end{equation}

Thus the eigenvalue $\Lambda \left( \overline{q} \right)$ of (\ref{eq:Lam_in_R}) has $R$ roots

\begin{equation} \label{eq:Lam_roots}
  \Lambda_k \left( \overline{q} \right) = \sqrt[R]{\lambda^Q} = \exp \left( \cfrac{Q * Ln \left( \lambda \right) + 2 * \pi * k * i}{R} \right), \ k \in [0, R-1],
\end{equation}

where $Ln \left( \lambda \right)$ is the principal value of the logarithm.

\subsection{The simplest cases}

Based on (\ref{eq:Lam_roots}), the simplest cases of R = 1 ... 6 are shown in Table \ref{tab:subspaces}, which has the following columns:

\begin{enumerate}
    \item R - the amount of circle-space points (see Figure \ref{fig:circle});
    \item WR - the amount of basis vectors $\overleftarrow{W}$ (see (\ref{eq:vec_W})) corresponding to $R$, which is equal to $2^R$;
    \item Q - the amount of quanta in the basis vector (see (\ref{eq:Q}));
    \item WQ - the amount of basis vectors $\overleftarrow{W}$ corresponding to $Q$, which is equal to $\displaystyle \binom{R}{Q} = \frac{R!}{Q!*(R-Q)!}$;
    \item $\displaystyle \overline{q}$ - the compound numbers of basis vectors $\displaystyle \overleftarrow{W_{\overline{q}}}$, for example, vector $\displaystyle \overleftarrow{W_{100}}$ has the compound number $100$;
    \item $\displaystyle \Lambda \left( \overline{q} \right)$ - the eigenvalues corresponding to $\displaystyle \overline{q}$ (see (\ref{eq:Lam_roots}));
    \item $A\Lambda$ - the amount of the eigenvalues corresponding to $\displaystyle \Lambda \left( \overline{q} \right)$, which is equal to the degree of the root.
\end{enumerate}

\begingroup
\setlength{\tabcolsep}{2pt}
\renewcommand{\arraystretch}{1.6}
\begin{table}[ht!]
\begin{center}
\begin{tabular}{| c | c | c | c | c | c | c |} 
\hline \hline
R & WR & Q & WQ & $\displaystyle \overline{q}$ & $\displaystyle \Lambda \left( \overline{q} \right)$ & $A\Lambda$ \\
\hline \hline
\multirow{2}{*}{1} & \multirow{2}{*}{2} & 0 & 1 & 0 & 1 & 1 \\
\cline{3-7}
& & 1 & 1 & 1 & $\lambda$ & 1 \\
\hline \hline
\multirow{3}{*}{2} & \multirow{3}{*}{4} & 0 & 1 & 00 & 1 & 1 \\
\cline{3-7}
& & 1 & 2 & 01, 10 & $\sqrt{\lambda}$ & 2 \\
\cline{3-7}
& & 2 & 1 & 11 & $\lambda$ & 1 \\
\hline \hline
\multirow{4}{*}{3} & \multirow{4}{*}{8} & 0 & 1 & 000 & 1 & 1 \\
\cline{3-7}
& & 1 & 3 & 001, 010, 100 & $\sqrt[3]{\lambda}$ & 3 \\
\cline{3-7}
& & 2 & 3 & 011, 110, 101 & $\sqrt[3]{\lambda^2}$ & 3 \\
\cline{3-7}
& & 3 & 1 & 111 & $\lambda$ & 1 \\
\hline \hline
\multirow{5}{*}{4} & \multirow{5}{*}{16} & 0 & 1 & 0000 & 1 & 1 \\
\cline{3-7}
& & 1 & 4 & 0001, 0010, 0100, 1000 & $\sqrt[4]{\lambda}$ & 4 \\
\cline{3-7}
& & 2 & 6 & 0011, 0110, 1100, 1001, 0101, 1010 & $\sqrt[4]{\lambda^2}$ & 4 \\
\cline{3-7}
& & 3 & 4 & 0111, 1110, 1101, 1011 & $\sqrt[4]{\lambda^3}$ & 4\\
\cline{3-7}
& & 4 & 1 & 1111 & $\lambda$ & 1 \\
\hline \hline
\multirow{6}{*}{5} & \multirow{6}{*}{32} & 0 & 1 & 00000 & 1 & 1 \\
\cline{3-7}
& & 1 & 5 & 00001, 00010, 00100, 01000, 10000 & $\sqrt[5]{\lambda}$ & 5 \\
\cline{3-7}
& & 2 & 10 & 00011, 00110, 01100, 11000, 10001, 
00101, 01010, 10100, 01001, 10010 & $\sqrt[5]{\lambda^2}$ & 5 \\
\cline{3-7}
& & 3 & 10 & 00111, 01110, 11100, 11001, 10011, 
01011, 10110, 01101, 11010, 10101 & $\sqrt[5]{\lambda^3}$ & 5\\
\cline{3-7}
& & 4 & 5 & 01111, 11110, 11101, 11011, 10111 & $\sqrt[5]{\lambda^4}$ & 5 \\
\cline{3-7}
& & 5 & 1 & 11111 & $\lambda$ & 1 \\
\hline \hline
\multirow{10}{*}{6} & \multirow{10}{*}{64} & 0 & 1 & 000000 & 1 & 1 \\
\cline{3-7}
& & 1 & 6 & 000001, 000010, 000100, 001000, 010000, 100000 & $\sqrt[6]{\lambda}$ & 6 \\
\cline{3-7}
& & \multirow{2}{*}{2} & \multirow{2}{*}{15} & 000011, 000110, 001100, 011000, 110000, 100001 & \multirow{2}{*}{$\displaystyle \sqrt[6]{\lambda^2}$} & \multirow{2}{*}{6} \\
& & & & 000101, 001010, 010100, 101000, 010001, 100010, 001001, 010010, 100100 & & \\
\cline{3-7}
& & \multirow{2}{*}{3} & \multirow{2}{*}{20} & 000111, 001110, 011100, 111000, 110001, 100011, 001101, 011010, 110100, 101001 & \multirow{2}{*}{$\displaystyle \sqrt[6]{\lambda^3}$} & \multirow{2}{*}{6} \\
& & & & 010011, 100110, 
001011, 010110, 101100, 011001, 110010, 100101, 010101, 101010 & & \\
\cline{3-7}
& & \multirow{2}{*}{4} & \multirow{2}{*}{15} & 001111, 011110, 111100, 111001, 110011, 100111 & \multirow{2}{*}{$\displaystyle \sqrt[6]{\lambda^4}$} & \multirow{2}{*}{6} \\
& & & & 010111, 101110, 011101, 111010, 110101, 101011, 011011, 110110, 101101 & & \\
\cline{3-7}
& & 5 & 6 & 011111, 111110, 111101, 111011, 110111, 101111 & $\sqrt[6]{\lambda^5}$ & 6 \\
\cline{3-7}
& & 6 & 1 & 111111 & $\lambda$ & 1 \\
\hline \hline
\end{tabular}
\end{center}
\caption{The division of state-space into subspaces according to the values of Q for the simplest cases of R = 1 ... 6}
\label{tab:subspaces}
\end{table}
\endgroup

\section{The eigenvectors of the elementary action} \label{sec:eigenvectors}

\subsection{Splitting the state-space subspaces into ordered sub-subspaces closed under elementary action} \label{subsec:sub_subspaces}

In the previous section, the state-space was divided into subspaces according to the values of Q of (\ref{eq:Q}). The subspaces may be splitted further into sub-subspaces and ordered with respect for the elementary action $\overleftrightarrow{Z}$ of (\ref{eq:mat_Z}) for $R > 1$ and $0 < Q < R$. To do this, one may use the following algorithm in each subspace:

\begin{enumerate}
    \item Any basis vector $\displaystyle W_{q_1 q_2 \ldots q_R}$ is selected and ordered at number 1. 
    \item The numbered 1 basis vector $\displaystyle W_{q_1 q_2 \ldots q_R}$ is acted upon by the elementary action $\overleftrightarrow{Z}$ of (\ref{eq:mat_Z}). Which is the product of the diagonal matrix ${\overleftrightarrow{\lambda}}_{[R]} \left( \lambda \right)$ of (\ref{eq:mat_Lam_R}) and the cyclic shift matrix $\overleftrightarrow{P_r}$ of (\ref{eq:mat_P_r}). The matrix $\overleftrightarrow{P_r}$ transforms the basis vector $\displaystyle W_{q_1 q_2 \ldots q_R}$ into the basis vector $\displaystyle W_{q_2 \ldots q_R q_1}$ according to (\ref{eq:mat_P_by_W}). And the matrix ${\overleftrightarrow{\lambda}}_{[R]} \left( \lambda \right)$ just multiplies the basis vector by the coefficient: 1 or $\lambda$, depending on the last subnumber. The transformed basis vector $\displaystyle W_{q_2 \ldots q_R q_1}$ has the same $Q$ according to (\ref{eq:Q}). So let it be ordered at number 2.
    \item The numbered 2 basis vector $\displaystyle W_{q_2 \ldots q_R q_1}$ is also acted upon by the elementary action $\overleftrightarrow{Z}$ of (\ref{eq:mat_Z}), etc. The action continues until vector numbered 1 is received, which does not exceed $R$ according to (\ref{eq:mat_Z_by_R}). Then the splitting off of the sub-subspace is completed.
    \item If unselected vectors remain, the splitting off of a new sub-subspace begins from step 1 of the algorithm.
\end{enumerate}

Based on the above algorithm, the simplest cases of R = 2 ... 6 are shown in Table \ref{tab:sub_subspaces}. In which, in comparison with Table \ref{tab:subspaces}, two columns are added: \textnumero - table row number, WS - the amount of vectors of the sub-subspace.

\begingroup
\setlength{\tabcolsep}{2pt}
\renewcommand{\arraystretch}{1.6}
\begin{table}[ht!]
\begin{center}
\begin{tabular}{| c | c | c | c | c | c | c | c |} 
\hline \hline
R & Q & WQ & \textnumero & $\displaystyle \overline{q}$ & WS & $\displaystyle \Lambda \left( \overline{q} \right)$ & $A\Lambda$ \\
\hline \hline
2 & 1 & 2 & 1 & 01, 10 & 2 & $\sqrt{\lambda}$ & 2 \\
\hline \hline
\multirow{2}{*}{3} & 1 & 3 & 2 & 001, 010, 100 & 3 & $\sqrt[3]{\lambda}$ & 3 \\
\cline{2-8}
& 2 & 3 & 3 & 011, 110, 101 & 3 & $\sqrt[3]{\lambda^2}$ & 3 \\
\hline \hline
\multirow{4}{*}{4} & 1 & 4 & 4 & 0001, 0010, 0100, 1000 & 4 & $\sqrt[4]{\lambda}$ & 4\\
\cline{2-8}
& \multirow{2}{*}{2} & \multirow{2}{*}{6} & 5 & 0011, 0110, 1100, 1001 & 4 & $\sqrt[4]{\lambda^2}$ & 4 \\
\cline{4-8}
& & & 6 & 0101, 1010 & 2 & $\sqrt{\lambda}$ & 2 \\
\cline{2-8}
& 3 & 4 & 7 & 0111, 1110, 1101, 1011 & 4 & $\sqrt[4]{\lambda^3}$ & 4\\
\hline \hline
\multirow{6}{*}{5} & 1 & 5 & 8 & 00001, 00010, 00100, 01000, 10000 & 5 & $\sqrt[5]{\lambda}$ & 5 \\
\cline{2-8}
& \multirow{2}{*}{2} & \multirow{2}{*}{10} & 9 & 00011, 00110, 01100, 11000, 10001 & 5 & \multirow{2}{*}{$\sqrt[5]{\lambda^2}$} & \multirow{2}{*}{5} \\
\cline{4-6}
& & & 10 & 00101, 01010, 10100, 01001, 10010 & 5 & & \\
\cline{2-8}
& \multirow{2}{*}{3} & \multirow{2}{*}{10} & 11 & 00111, 01110, 11100, 11001, 10011 & 5 & \multirow{2}{*}{$\sqrt[5]{\lambda^3}$} & \multirow{2}{*}{5} \\
\cline{4-6}
& & & 12 & 01011, 10110, 01101, 11010, 10101 & 5 & & \\
\cline{2-8}
& 4 & 5 & 13 & 01111, 11110, 11101, 11011, 10111 & 5 & $\sqrt[5]{\lambda^4}$ & 5 \\
\hline \hline
\multirow{12}{*}{6} & 1 & 6 & 14 & 000001, 000010, 000100, 001000, 010000, 100000 & 6 & $\sqrt[6]{\lambda}$ & 6 \\
\cline{2-8}
& \multirow{3}{*}{2} & \multirow{3}{*}{15} & 15 & 000011, 000110, 001100, 011000, 110000, 100001 & 6 & \multirow{2}{*}{$\sqrt[6]{\lambda^2}$} & \multirow{2}{*}{6} \\
\cline{4-6}
& & & 16 & 000101, 001010, 010100, 101000, 010001, 100010 & 6 & & \\
\cline{4-8}
& & & 17 & 001001, 010010, 100100 & 3 & $\sqrt[3]{\lambda}$ & 3 \\
\cline{2-8}
& \multirow{4}{*}{3} & \multirow{4}{*}{20} & 18 & 000111, 001110, 011100, 111000, 110001, 100011 & 6 & \multirow{3}{*}{$\sqrt[6]{\lambda^3}$} & \multirow{3}{*}{6} \\
\cline{4-6}
& & & 19 & 001101, 011010, 110100, 101001, 010011, 100110 & 6 & & \\
\cline{4-6}
& & & 20 & 001011, 010110, 101100, 011001, 110010, 100101 & 6 & & \\
\cline{4-8}
& & & 21 & 010101, 101010 & 2 & $\sqrt{\lambda}$ & 2 \\
\cline{2-8}
& \multirow{3}{*}{4} & \multirow{3}{*}{15} & 22 & 001111, 011110, 111100, 111001, 110011, 100111 & 6 & \multirow{2}{*}{$\sqrt[6]{\lambda^4}$} & \multirow{2}{*}{6} \\
\cline{4-6}
& & & 23 & 010111, 101110, 011101, 111010, 110101, 101011 & 6 & & \\
\cline{4-8}
& & & 24 & 011011, 110110, 101101 & 3 & $\sqrt[3]{\lambda^2}$ & 3 \\
\cline{2-8}
& 5 & 6 & 25 & 011111, 111110, 111101, 111011, 110111, 101111 & 6 & $\sqrt[6]{\lambda^5}$ & 6 \\
\hline \hline
\end{tabular}
\end{center}
\caption{The splitting of the state-space subspaces into sub-subspaces for the simplest cases of R = 2 ... 6}
\label{tab:sub_subspaces}
\end{table}
\endgroup

\subsection{Determining eigenvectors using a simple example}

Let us determine eigenvectors for sub-subspace \textnumero 5 (see Table \ref{tab:sub_subspaces}) having: the amount of circle-space points $R=4$, the amount of quanta $Q=2$, four basis vectors $\{ \overleftarrow{W_{0011}}, \ \overleftarrow{W_{0110}}, \ \overleftarrow{W_{1100}}, \ \overleftarrow{W_{1001}} \}$, four eigenvalues $\sqrt[4]{\lambda^2}$.

For some eigenvalue $\Lambda_k$ of (\ref{eq:Lam_roots}), the eigenvector $\overleftarrow{e_k}$ is the linear combination of its state-space coordinates with respect for the four basis vectors.

\begin{equation} \label{eq:vec_e}
    \overleftarrow{e_k} = f_k \left( 0 \right) * \overleftarrow{W_{0011}} + f_k \left( \Delta u \right) * \overleftarrow{W_{0110}} + f_k \left( 2 * \Delta u \right) * \overleftarrow{W_{1100}} + f_k \left( 3 * \Delta u \right) * \overleftarrow{W_{1001}},
\end{equation}

where the sought-for state-space coordinates $f_k$ may be interpreted as follows: $f_k \left( 0 \right)$ describes the location of the quanta configuration 0011 at the initial circle-space point $u_1$ having circle-space coordinate 0 (see Figure \ref{fig:circle}), $f_k \left( \Delta u \right)$ describes the location of the same quanta configuration at the next circle-space point $u_R$ having circle-space coordinate $\Delta u$, etc.

The characteristic equation for the elementary action $\overleftrightarrow{Z}$ of (\ref{eq:mat_Z}) is

\begin{equation} \label{eq:char_eq}
    \overleftrightarrow{Z} * \overleftarrow{e_k} = {\overleftrightarrow{\lambda}}_{[4]} \left( \lambda \right) * \overleftrightarrow{P_r} * \overleftarrow{e_k} = \Lambda_k * \overleftarrow{e_k}.
\end{equation}

Let the subnumbers of the first basis vector (which are now 0011) be written as a function $\phi$ of points in circle-space

\begin{equation} \label{eq:phi}
    \phi \left( 0 \right) = 0, \ \phi \left( \Delta u \right) = 0,  \ \phi \left( 2 * \Delta u \right) = 1,  \ \phi \left( 3 * \Delta u \right) = 1,
\end{equation}

where (\ref{eq:Q}) gives

\begin{equation} \label{eq:sum_phi}
    \phi \left( 0 \right) + \phi \left( \Delta u \right) + \phi \left( 2 * \Delta u \right) + \phi \left( 3 * \Delta u \right) = Q.
\end{equation}

Accounting (\ref{eq:mat_P_by_W}) and (\ref{eq:mat_Lam_by_W}) with (\ref{eq:phi}) gives

\begin{equation} \label{eq:mat_Z_by_W_with_phi}
\begin{aligned}
    &\overleftrightarrow{Z} * \overleftarrow{W_{0011}} = {\overleftrightarrow{\lambda}}_{[4]} \left( \lambda \right) * \overleftarrow{W_{0110}} = \lambda^{\phi \left( 0 \right)} * \overleftarrow{W_{0110}}, \\
    &\overleftrightarrow{Z} * \overleftarrow{W_{0110}} = {\overleftrightarrow{\lambda}}_{[4]} \left( \lambda \right) * \overleftarrow{W_{1100}} = \lambda^{\phi \left( \Delta u \right)} * \overleftarrow{W_{1100}}, \\
    &\overleftrightarrow{Z} * \overleftarrow{W_{1100}} = {\overleftrightarrow{\lambda}}_{[4]} \left( \lambda \right) * \overleftarrow{W_{1001}} = \lambda^{\phi \left( 2 * \Delta u \right)} * \overleftarrow{W_{1001}}, \\
    &\overleftrightarrow{Z} * \overleftarrow{W_{1001}} = {\overleftrightarrow{\lambda}}_{[4]} \left( \lambda \right) * \overleftarrow{W_{0011}} = \lambda^{\phi \left( 3 * \Delta u \right)} * \overleftarrow{W_{0011}}.
\end{aligned}
\end{equation}

Substitution of (\ref{eq:vec_e}) into (\ref{eq:char_eq}) taking into account (\ref{eq:mat_Z_by_W_with_phi}) gives

\begin{equation} \label{eq:char_eq_is_0}
    \begin{aligned}
    &\left( \lambda^{\phi \left( 0 \right)} * f_k \left( 0 \right) - \Lambda_k * f_k \left( \Delta u \right) \right) * \overleftarrow{W_{0110}} \\
    + &\left( \lambda^{\phi \left( \Delta u \right)} * f_k \left( \Delta u \right) - \Lambda_k * f_k \left( 2 * \Delta u \right) \right) * \overleftarrow{W_{1100}} \\
    + &\left( \lambda^{\phi \left( 2 * \Delta u \right)} * f_k \left( 2 * \Delta u \right) - \Lambda_k * f_k \left( 3 * \Delta u \right) \right) * \overleftarrow{W_{1001}} \\
    + &\left( \lambda^{\phi \left( 3 * \Delta u \right)} * f_k \left( 3 * \Delta u \right) - \Lambda_k * f_k \left( 0 \right) \right) * \overleftarrow{W_{0011}} = 0.
    \end{aligned}
\end{equation}

The coefficients for the basis vectors must be equal to zero. Assuming the initial coordinate $f_k \left( 0 \right) = 1$, the first three coefficients determine the other coordinates, and the fourth one should naturally equal zero

\begin{equation} \label{eq:fs_got}
\begin{aligned}
    f_k \left( 0 \right) &= 1, \\
    f_k \left( \Delta u \right) &= {\Lambda_k}^{-1} * \lambda^{\phi \left( 0 \right)} * f_k \left( 0 \right) = {\Lambda_k}^{-1} * \lambda^{\phi \left( 0 \right)} = {\Lambda_k}^{-1}, \\
    f_k \left( 2 * \Delta u \right) &= {\Lambda_k}^{-1} * \lambda^{\phi \left( \Delta u \right)} * f_k \left( \Delta u \right) = {\Lambda_k}^{-2} * \lambda^{\phi \left( 0 \right) + \phi \left( \Delta u \right)} = {\Lambda_k}^{-2}, \\
    f_k \left( 3 * \Delta u \right) &= {\Lambda_k}^{-1} * \lambda^{\phi \left( 2 * \Delta u \right)} * f_k \left( 2 * \Delta u \right) = {\Lambda_k}^{-3} * \lambda^{ \phi \left( 0 \right) + \phi \left( \Delta u \right) + \phi \left( 2 * \Delta u \right)} = \lambda * {\Lambda_k}^{-3}, \\
     1 &= {\Lambda_k}^{-1} * \lambda^{\phi \left( 3 * \Delta u \right)} * f_k \left( 3 * \Delta u \right) = {\Lambda_k}^{-4} * \lambda^{ \phi \left( 0 \right) + \phi \left( \Delta u \right) + \phi \left( 2 * \Delta u \right) + \phi \left( 3 * \Delta u \right)}.
\end{aligned}
\end{equation}

The last equation of (\ref{eq:fs_got}) is true due to (\ref{eq:sum_phi}) and (\ref{eq:Lam_in_R}).

Let the numerical calculations be performed for $\lambda$ equal to $\lambda_{21}$ of (\ref{eq:lambda}). Then $\lambda$ and eigenvalues $\Lambda_k$ of (\ref{eq:Lam_roots}) are: 

\begin{equation} \label{eq:lam_num}
    \lambda = \lambda_{21} = \cfrac{9-2\sqrt{73}}{9+2\sqrt{73}} \approx -0.3100, \ 
    \Lambda_0 = \sqrt[4]{\lambda^2} \approx 0.5568, \ 
    \Lambda_1 \approx i * 0.5568, \ 
    \Lambda_2 \approx -0.5568, \ 
    \Lambda_3 \approx -i * 0.5568.
\end{equation}

For $R = 4$, the eigenvectors have 16 elements each, numbered from 0 to $1111_2 = 15_{10}$. Their nonzero elements are calculated with (\ref{eq:fs_got}), taking into account (\ref{eq:phi}) and (\ref{eq:lam_num}). The calculation results are shown in Table \ref{tab:eigenvectors_N_5}, which has the following columns:

\begin{itemize}
    \item $n_2$ and $n_{10}$ - nonzero element number: $n_2$ - binary and $n_{10}$ - decimal;
    \item $\overleftarrow{e_0}, \ \overleftarrow{e_1}, \ \overleftarrow{e_2}, \ \overleftarrow{e_3}$ - eigenvectors.
\end{itemize}

The last table row $\Lambda$ shows the eigenvalues.

\begingroup
\setlength{\tabcolsep}{2pt}
\renewcommand{\arraystretch}{1.6}
\begin{table}[ht!]
\begin{center}
\begin{tabular}{| c | c || c | c | c | c |} 
\hline \hline
$n_2$ & $n_{10}$ & $\overleftarrow{e_0}$ & $\overleftarrow{e_1}$ & $\overleftarrow{e_2}$ & $\overleftarrow{e_3}$ \\
\hline \hline
0011 & 3 
& $f_0 \left( 0 \right) = 1$ 
& $f_1 \left( 0 \right) = 1$ 
& $f_2 \left( 0 \right) = 1$ 
& $f_3 \left( 0 \right) = 1$ \\
\hline
\multirow{2}{*}{0110} & \multirow{2}{*}{6} 
& $f_0 \left( \Delta u \right) = {\Lambda_0}^{-1}$ 
& $f_1 \left( \Delta u \right) = {\Lambda_1}^{-1}$ 
& $f_2 \left( \Delta u \right) = {\Lambda_2}^{-1}$ 
& $f_3 \left( \Delta u \right) = {\Lambda_3}^{-1}$ 
\\ & & $\approx 1.796$ & $\approx -i*1.796$ & $\approx -1.796$ & $\approx i*1.796$ \\
\hline
\multirow{2}{*}{1001} & \multirow{2}{*}{9} 
& $f_0 \left( 3 * \Delta u \right) = \lambda * {\Lambda_0}^{-3}$ 
& $f_1 \left( 3 * \Delta u \right) = \lambda * {\Lambda_1}^{-3}$ 
& $f_2 \left( 3 * \Delta u \right) = \lambda * {\Lambda_2}^{-3}$ 
& $f_3 \left( 3 * \Delta u \right) = \lambda * {\Lambda_3}^{-3}$ 
\\ & & $\approx -1.796$ & $\approx -i*1.796$ & $\approx 1.796$ & $\approx i*1.796$ \\ 
\hline
\multirow{2}{*}{1100} & \multirow{2}{*}{12} 
& $f_0 \left( 2 * \Delta u \right) = {\Lambda_0}^{-2}$ 
& $f_1 \left( 2 * \Delta u \right) = {\Lambda_1}^{-2}$ 
& $f_2 \left( 2 * \Delta u \right) = {\Lambda_2}^{-2}$ 
& $f_3 \left( 2 * \Delta u \right) = {\Lambda_3}^{-2}$ 
\\ & & $\approx 3.226$ & $\approx -3.226$ & $\approx 3.226$ & $\approx -3.226$ \\
\hline \hline
\multicolumn{2}{|c||}{\multirow{2}{*}{$\Lambda$}} & $\Lambda_0 = \sqrt[4]{\lambda^2}$ 
& $\Lambda_1 = i*\sqrt[4]{\lambda^2}$ & $\Lambda_2 = -\sqrt[4]{\lambda^2}$ 
& $\Lambda_3 =-i* \sqrt[4]{\lambda^2}$ 
\\ \multicolumn{2}{|c||}{} & $\approx 0.5568$ & $\approx i*0.5568$ & $\approx -0.5568$ 
&  $\approx -i*0.5568$ \\
\hline \hline
\end{tabular}
\end{center}
\caption{The nonzero eigenvectors elements for $R=4$ sub-subspace \textnumero 5 of Table \ref{tab:sub_subspaces}}
\label{tab:eigenvectors_N_5}
\end{table}
\endgroup

\subsection{Generalization of determining eigenvectors}

In Subsection \ref{subsec:eigenval_gen}, the state-space was divided into the subspaces according to the amount of quanta $Q$ in the basis vector $\overleftarrow{W_{q_1 q_2 \ldots q_R}}$ (see Table \ref{tab:subspaces}). In particular, for any $R$ at $Q=0$ and $Q=R$ there were singleton subspaces having maximum eigenvalue and minimum absolute value eigenvalue respectively. 

In Subsection \ref{subsec:sub_subspaces}, at $R>1$ and $0<Q<R$, the subspaces were further splitted into ordered sub-subspaces closed under elementary action (see Table \ref{tab:sub_subspaces}) having their dimentions at most $R$. In particular, for any $R>1$ at $Q=1$ and $Q=R-1$ there was one R-dimensional subspace which was not splitted and just ordered. And for $R>3$ at $1<Q<R-1$, the subspaces were always splitted as the amount of their basis vectors $\displaystyle \binom{R}{Q}$ exceeded $R$.

When $R$ and $Q$ have a common divisor $d_{R,Q}$, then $R/d_{R,Q}$-dimensional subspaces appear, similar to $R/d_{R,Q}$-dimensional subspaces at $Q/d_{R,Q}$. See in Table \ref{tab:sub_subspaces}: sub-subspaces \textnumero 6 for $R=4$ at $Q=2$ and \textnumero 21 for $R=6$ at $Q=3$ are similar to sub-subspace \textnumero 1 for $R=2$ at $Q=1$, sub-subspace \textnumero 17 for $R=6$ at $Q=2$ is similar to sub-subspace \textnumero 2 for $R=3$ at $Q=1$, sub-subspace \textnumero 24 for $R=6$ at $Q=4$ is similar to sub-subspace \textnumero 3 for $R=3$ at $Q=2$.

Thus, when $R$ is a prime number, its sub-subspaces at any $1<Q<R-1$ are only R-dimensional (see in Table 2: R = 2, 3, 5).

Now let the generalization of determining eigenvectors be performed on the base of the previous subsection for some ordered R-dimensional sub-subspace having any $R>1, \ 0<Q<R$, the initial basis vector $\overleftarrow{W_{q_1 q_2 \ldots q_R}}$, $R$ eigenvalues $\Lambda_k$ of (\ref{eq:Lam_roots}) for $\ k \in [0, R-1]$.

The function $\phi \left( u \right)$ of (\ref{eq:phi}) can be generalized to

\begin{equation} \label{eq:phi_gen}
    \phi \left( 0 \right) = q_1, \ \phi \left( \Delta u \right) = q_2,  \ \ldots,  \ \phi \left( \left( R - 1 \right) * \Delta u \right) = q_R.
\end{equation}

Let the function $\Phi \left( u \right)$ be introduced based on (\ref{eq:phi_gen})

\begin{equation} \label{eq:Phi_gen}
\begin{aligned}
    &\Phi \left( 0 \right) = 0, \ 
    \Phi \left( \Delta u \right) = \phi \left( 0 \right), \ 
    \Phi \left( 2 * \Delta u \right) =  \phi \left( 0 \right) + \phi \left( \Delta u \right),  \ \ldots,  \\ 
    &\Phi \left( \left( R -1 \right) * \Delta u \right) = \phi \left( 0 \right) + \phi \left( \Delta u \right) + \ldots + \phi \left( \left( R - 2 \right) * \Delta u \right).
\end{aligned}
\end{equation}

Then the eigenvector $\overleftarrow{e_k}$ coordinates of (\ref{eq:fs_got}) can be generalized to

\begin{equation} \label{eq:fs_gen}
\begin{aligned} 
    f_k \left( 0 \right) &= 1, \\
    f_k \left( \Delta u \right) &= {\Lambda_k}^{-1} * \lambda^{\Phi \left( \Delta u \right)}, \\
    f_k \left( 2 * \Delta u \right) &= {\Lambda_k}^{-2} * \lambda^{\Phi \left( 2 * \Delta u \right)}, \\ \ldots \\
    f_k \left(  \left( R -1 \right) * \Delta u \right) &= {\Lambda_k}^{-\left( R -1 \right)} * \lambda^{ \Phi \left( \left( R -1 \right) * \Delta u \right)}, \\
\end{aligned}
\end{equation}

or finally

\begin{equation} \label{eq:fs_gen_gen}
    f_k \left( u \right) = \lambda^{\Phi \left( u \right)} * \exp{ \left( -\cfrac{u}{\Delta u} * \ln{ \left( \Lambda_k \right) } \right) } .
\end{equation}

Let the eigenvector $\overleftarrow{e_k}$ with coordinates $f_k \left( u \right)$ of (\ref{eq:fs_gen_gen}) undergo several elementary actions $\overleftrightarrow{Z}$ of (\ref{eq:mat_Z}). One elementary action on its eigenvector leads to the multiplication of its coefficients by the eigenvalue $\Lambda_k$, which was called the motion per time step $\Delta t$ in the simplest quantum field model (see Subsection \ref{subsec:QuantumField}). Then several elementary actions may be called the motion per time $t$, where the amount of the elementary actions $t/ \Delta t$ is a natural number. Thus, for the coordinates of the eigenvector at time $t$, the wave function $F_k \left( u, t \right)$ may be introduced as follows

\begin{equation} \label{eq:wave_func}
    F_k \left( u, t \right) = f_k \left( u \right) * \exp{ \left( \cfrac{t}{\Delta t} * \ln{ \left( \Lambda_k \right) } \right) } .
\end{equation}

Let the following universal values be introduced: the uncertainty $\delta$, as well as the frequency step $\Delta \omega$ and the wavenumber step $\Delta \kappa$

\begin{equation} \label{eq:uncert}
    \delta = \cfrac{2 * \pi}{R}, \quad 
    \Delta \omega * \Delta t = \delta, \quad 
    \Delta \kappa * \Delta u = \delta.
\end{equation}

Substituting $f_k \left( u \right)$ of (\ref{eq:fs_gen_gen}) and $\Lambda_k$ of (\ref{eq:Lam_roots}) to (\ref{eq:wave_func}), taking into account $(\ref{eq:uncert})$, gives the wave function $F_k \left( u, t \right)$ in the final form

\begin{equation} \label{eq:wave_func_fin}
    F_k \left( u, t \right) = \lambda^{\Phi \left( u \right)} * \exp{ \left( \Omega_k * t - K_k * u \right) } ,
\end{equation}

where the full frequency $\Omega_k$ and the full wavenumber $K_k$ are

\begin{equation} \label{eq:full_wave_vals}
    \Omega_k = \cfrac{ Q * Ln \left( \lambda \right) }{R * \Delta t} + i * k * \Delta \omega , \quad
    K_k = \cfrac{Q * Ln \left( \lambda \right) }{R * \Delta u} + i * k * \Delta \kappa. 
\end{equation}

Note that $\Omega_k$ is directly proportional to $K_k$ through the universal value c

\begin{equation} \label{eq:c}
    \Omega_k = c * K_k, \quad c = \cfrac{\Delta u}{\Delta t}.
\end{equation}

\subsection{A little fun} \label{subsec:fun}

For real positive $0 < \lambda < 1$, there is real negative $Ln \left( \lambda \right) < 0$. Then (\ref{eq:wave_func_fin}) and (\ref{eq:full_wave_vals}) describe damped waves. This entails various wave effects, some of which may be interpreted funnily:

\begin{enumerate}
    \item The group velocity of any wave function $F_k \left( u, t \right)$ of (\ref{eq:wave_func_fin}) is equal to its phase velocity, which is equal to universal value $c$ of (\ref{eq:c}). This may be interpreted as the simplest model of the special relativity.
    \item The wave functions propagate independently in homogeneous space-time. But when there is heterogeneity, interaction arises between them. This may be interpreted as the simplest model of the general relativity.
    \item (\ref{eq:uncert}) may be interpreted as the simplest model of the uncertainty principle.
    \item The basis vectors $\overleftarrow{W_{\overline{q}}}$ of (\ref{eq:vec_W}) are not the eigenvectors of the elementary action $\overleftrightarrow{Z}$ of (\ref{eq:mat_Z}) (excluding the vacuum eigenvector $\overleftarrow{W_{00 \ldots 0}}$ of (\ref{eq:vacc}) and the full eigenvector $\overleftarrow{W_{11 \ldots 1}}$ of (\ref{eq:filled})), so they behave like waves. However, these same basis vectors $\overleftarrow{W_{\overline{q}}}$ are the eigenvectors of the elementary action $\overleftrightarrow{Z}$ powered $R$ (see (\ref{eq:mat_Z_by_R})), so they behave like particles. This may be interpreted as the simplest model of the wave-particle duality.
\end{enumerate}

\section{The Jordan normal form of the elementary action} \label{sec:Jordan}

\subsection{Determining the transformation matrix using a simple example}

Let us consider a simple example: the amount of circle-space points is $R=4$. Then the elementary action $\overleftrightarrow{Z} \left( \lambda \right)$ of (\ref{eq:mat_Z}) has $2^4\times2^4 = 16\times16$ matrix, let its rows and columns be numbered from $0000_2=0_{10}$ to $1111_2=15_{10}$. Let 16 basis vectors be $\overleftarrow{W_{0000}}, \ldots, \overleftarrow{W_{1111}}$ of (\ref{eq:vec_W}). Let the numerical calculations be performed for $\lambda$ equal to $\lambda_{21}$ of (\ref{eq:lambda}).

The Jordan normal form is obtained by some similarity transformation ${\overleftrightarrow{S_6}}^{-1} * \overleftrightarrow{Z} \left( \lambda \right) * \overleftrightarrow{S_6}$. Let the transformation matrix $\overleftrightarrow{S_6}$ be built of 16 column eigenvectors of the elementary action, the coordinates of which $f_k \left( u \right)$ are calculated by (\ref{eq:fs_gen_gen}). Further, only nonzero transformation matrix elements will be shown.

\begin{enumerate}
\setcounter{enumi}{-1}
    \item The vacuum basis vector $\overleftarrow{W_{0000}}$ is the eigenvector with the eigenvalue equal to 1 (see (\ref{eq:vacc})). Let it be the transformation matrix column numbered $0000_2=0_{10}$, so this column has a single non-zero element at the top equal to 1.
    
    Similarly, the full basis vector $\overleftarrow{W_{1111}}$ is the eigenvector with the eigenvalue equal to $\lambda$ (see (\ref{eq:filled})). Let it be the transformation matrix column numbered $1111_2=15_{10}$, so this column has a single non-zero element at the bottom equal to 1.
    
    This is shown in Table \ref{tab:col_N_0_15}, where $n_2$ and $n_{10}$ are binary and decimal row/column number, row $\Lambda$ is the eigenvalue.
    
\begingroup
\setlength{\tabcolsep}{2pt}
\renewcommand{\arraystretch}{1.6}
\begin{table}[ht!]
\begin{center}
\begin{tabular}{| c | c || c |} 
\hline \hline
$n_2$ & & 0 \\
\hline
& $n_{10}$ & 0 \\
\hline \hline
0 & 0 & 1 \\
\hline \hline
\multicolumn{2}{|c||}{$\Lambda$} & 1 \\
\hline \hline
\end{tabular}
\hspace{2cm}
\begin{tabular}{| c | c || c |} 
\hline \hline
$n_2$ & & 1111 \\
\hline
& $n_{10}$ & 15 \\
\hline \hline
1111 & 15 & 1 \\
\hline \hline
\multicolumn{2}{|c||}{$\Lambda$} & $\lambda \approx -0.3100$ \\
\hline \hline
\end{tabular}
\end{center}
\caption{The nonzero transformation matrix elements of columns 0 and 15}
\label{tab:col_N_0_15}
\end{table}
\endgroup

    \item The sub-subspace \textnumero 4 of Table \ref{tab:sub_subspaces} has dimension 4, let its eigenvectors form the transformation matrix columns numbered $0001_2 = 1_{10}, \ 0010_2 = 2_{10}, \ 0011_2 = 3_{10}, \ 0100_2 = 4_{10}$. The first basis vector is $\overleftarrow{W_{0001}}$, so the functions $\phi \left( u \right)$ of (\ref{eq:phi_gen}) and $\Phi \left( u \right)$ of (\ref{eq:Phi_gen}) are 
    
\begin{equation*}
\begin{aligned}
    &\phi \left( 0 \right) = 0, \ \phi \left( \Delta u \right) = 0,  \ \phi \left( 2 * \Delta u \right) = 0,  \ \phi \left( 3 * \Delta u \right) = 1; \\
    &\Phi \left( 0 \right) = 0, \ \Phi \left( \Delta u \right) = 0, \ \Phi \left( 2 * \Delta u \right) = 0, \ \Phi \left( 3 * \Delta u \right) =  0.
\end{aligned}
\end{equation*}

The calculation results for the eigenvector coordinates $f_k \left( u \right)$ of (\ref{eq:fs_gen_gen}) are shown in Table \ref{tab:col_N_1_to_5}.

\begingroup
\setlength{\tabcolsep}{2pt}
\renewcommand{\arraystretch}{1.6}
\begin{table}[ht!]
\begin{center}
\begin{tabular}{| c | c || c | c | c | c |} 
\hline \hline
$n_2$ & & 0001 & 0010 & 0011 & 0100 \\
\hline
& $n_{10}$ & 1 & 2 & 3 & 4 \\
\hline \hline
0001 & 1 
& $f_0 \left( 0 \right) = 1$ 
& $f_1 \left( 0 \right) = 1$ 
& $f_2 \left( 0 \right) = 1$ 
& $f_3 \left( 0 \right) = 1$ \\
\hline
\multirow{2}{*}{0010} & \multirow{2}{*}{2} 
& $f_0 \left( \Delta u \right) = {\Lambda_0}^{-1}$ 
& $f_1 \left( \Delta u \right) = {\Lambda_1}^{-1}$
& $f_2 \left( \Delta u \right) = {\Lambda_2}^{-1}$ 
& $f_3 \left( \Delta u \right) = {\Lambda_3}^{-1}$ 
\\ & & $\approx 0.9476-i*0.9476$ & $\approx -0.9476-i*0.9476$ & $\approx -0.9476+i*0.9476$ & $\approx 0.9476+i*0.9476$ \\
\hline
\multirow{2}{*}{0100} & \multirow{2}{*}{4} 
& $f_0 \left( 2 * \Delta u \right) = {\Lambda_0}^{-2}$ 
& $f_1 \left( 2 * \Delta u \right) = {\Lambda_1}^{-2}$ 
& $f_2 \left( 2 * \Delta u \right) = {\Lambda_2}^{-2}$ 
& $f_3 \left( 2 * \Delta u \right) = {\Lambda_3}^{-2}$ 
\\ & & $\approx -i*1.796$ & $\approx i*1.796$ & $\approx -i*1.796$ & $\approx i*1.796$ \\ 
\hline
\multirow{2}{*}{1000} & \multirow{2}{*}{8} 
& $f_0 \left( 3 * \Delta u \right) = {\Lambda_0}^{-3}$ 
& $f_1 \left( 3 * \Delta u \right) = {\Lambda_1}^{-3}$
& $f_2 \left( 3 * \Delta u \right) = {\Lambda_2}^{-3}$ 
& $f_3 \left( 3 * \Delta u \right) = {\Lambda_3}^{-3}$ 
\\ & & $\approx -1.702-i*1.702$ & $\approx 1.702-i*1.702$ & $\approx 1.702+i*1.702$ & $\approx -1.702+i*1.702$ \\
\hline \hline
\multicolumn{2}{|c||}{\multirow{2}{*}{$\Lambda$}} & $\Lambda_0 = \sqrt[4]{\lambda}$ 
& $\Lambda_1 = i*\sqrt[4]{\lambda}$ & $\Lambda_2 = -\sqrt[4]{\lambda}$ 
& $\Lambda_3 =-i* \sqrt[4]{\lambda}$ 
\\ \multicolumn{2}{|c||}{} & $\approx 0.5276+i*0.5276$ & $\approx -0.5276+i*0.5276$ & $\approx -0.5276-i*0.5276$ &  $\approx 0.5276-i*0.5276$ \\
\hline \hline
\end{tabular}
\end{center}
\caption{The nonzero transformation matrix elements of columns from 1 to 4}
\label{tab:col_N_1_to_5}
\end{table}
\endgroup

   \item The sub-subspace \textnumero 5 of Table \ref{tab:sub_subspaces} is calculated in Table \ref{tab:eigenvectors_N_5}. Let its 4 columns be numbered $0101_2 = 5_{10}, \ 0110_2 = 6_{10}, \ 0111_2 = 7_{10}, \ 1000_2 = 8_{10}$.

    \item The sub-subspace \textnumero 6 of Table \ref{tab:sub_subspaces} has dimension 2, so the calculation is similar to the two-dimensional sub-subspace \textnumero 1 of Table \ref{tab:sub_subspaces}. Let sub-subspace \textnumero 6 eigenvectors form the transformation matrix columns numbered $1001_2 = 9_{10}, \ 1010_2 = 10_{10}$. The first basis vector is $\overleftarrow{W_{0101}}$, so the functions $\phi \left( u \right)$ of (\ref{eq:phi_gen}) and $\Phi \left( u \right)$ of (\ref{eq:Phi_gen}) are 

\begin{equation*}
    \phi \left( 0 \right) = 0, \ \phi \left( \Delta u \right) = 1; \quad 
    \Phi \left( 0 \right) = 0, \ \Phi \left( \Delta u \right) = 0.
\end{equation*}

The calculation results for the eigenvector coordinates $f_k \left( u \right)$ of (\ref{eq:fs_gen_gen}) are shown in Table \ref{tab:col_N_9_10}.

\begingroup
\setlength{\tabcolsep}{2pt}
\renewcommand{\arraystretch}{1.6}
\begin{table}[ht!]
\begin{center}
\begin{tabular}{| c | c || c | c | c | c |} 
\hline \hline
$n_2$ & & 1001 & 1010 \\
\hline
& $n_{10}$ & 9 & 10 \\
\hline \hline
0101 & 5 
& $f_0 \left( 0 \right) = 1$ 
& $f_1 \left( 0 \right) = 1$ \\
\hline
1010 & 10 
& $f_0 \left( \Delta u \right) = {\Lambda_0}^{-1} \approx -i*1.796$ 
& $f_1 \left( \Delta u \right) = {\Lambda_1}^{-1} \approx i*1.796$ \\
\hline \hline
\multicolumn{2}{|c||}{$\Lambda$} & $\Lambda_0 = \sqrt{\lambda} \approx i*0.5568$ 
& $\Lambda_1 = -\sqrt{\lambda} \approx -i*0.5568$ \\
\hline \hline
\end{tabular}
\end{center}
\caption{The nonzero transformation matrix elements of columns 9 and 10}
\label{tab:col_N_9_10}
\end{table}
\endgroup

    \item The sub-subspace \textnumero 7 of Table \ref{tab:sub_subspaces} has dimension 4, let its eigenvectors form the transformation matrix columns numbered $1011_2 = 11_{10}, \ 1100_2 = 12_{10}, \ 1101_2 = 13_{10}, \ 1110_2 = 14_{10}$. The first basis vector is $\overleftarrow{W_{0111}}$, so the functions $\phi \left( u \right)$ of (\ref{eq:phi_gen}) and $\Phi \left( u \right)$ of (\ref{eq:Phi_gen}) are 
    
\begin{equation*}
\begin{aligned}
    &\phi \left( 0 \right) = 0, \ \phi \left( \Delta u \right) = 1,  \ \phi \left( 2 * \Delta u \right) = 1,  \ \phi \left( 3 * \Delta u \right) = 1; \\
    &\Phi \left( 0 \right) = 0, \ \Phi \left( \Delta u \right) = 0, \ \Phi \left( 2 * \Delta u \right) = 1, \ \Phi \left( 3 * \Delta u \right) =  2.
\end{aligned}
\end{equation*}

The calculation results for the eigenvector coordinates $f_k \left( u \right)$ of (\ref{eq:fs_gen_gen}) are shown in Table \ref{tab:col_N_11_to_14}.

\begingroup
\setlength{\tabcolsep}{2pt}
\renewcommand{\arraystretch}{1.6}
\begin{table}[ht!]
\begin{center}
\begin{tabular}{| c | c || c | c | c | c |} 
\hline \hline
$n_2$ & & 1011 & 1100 & 1101 & 1110 \\
\hline
& $n_{10}$ & 11 & 12 & 13 & 14 \\
\hline \hline
0111 & 7 
& $f_0 \left( 0 \right) = 1$ 
& $f_1 \left( 0 \right) = 1$ 
& $f_2 \left( 0 \right) = 1$ 
& $f_3 \left( 0 \right) = 1$ \\
\hline
\multirow{2}{*}{1011} & \multirow{2}{*}{11} 
& $f_0 \left( 3 * \Delta u \right) = {\lambda}^2 / {\Lambda_0}^3$ 
& $f_1 \left( 3 * \Delta u \right) = {\lambda}^2 / {\Lambda_1}^3$
& $f_2 \left( 3 * \Delta u \right) = {\lambda}^2 / {\Lambda_2}^3$ 
& $f_3 \left( 3 * \Delta u \right) = {\lambda}^2 / {\Lambda_3}^3$ 
\\ & & $\approx -0.9476-i*0.9476$ & $\approx 0.9476-i*0.9476$ & $\approx 0.9476+i*0.9476$ & $\approx -0.9476+i*0.9476$ \\
\hline
\multirow{2}{*}{1101} & \multirow{2}{*}{13} 
& $f_0 \left( 2 * \Delta u \right) = \lambda / {\Lambda_0}^2$ 
& $f_1 \left( 2 * \Delta u \right) = \lambda / {\Lambda_1}^2$ 
& $f_2 \left( 2 * \Delta u \right) = \lambda / {\Lambda_2}^2$ 
& $f_3 \left( 2 * \Delta u \right) = \lambda / {\Lambda_3}^2$ 
\\ & & $\approx i*1.796$ & $\approx -i*1.796$ & $\approx i*1.796$ & $\approx -i*1.796$ \\ 
\hline
\multirow{2}{*}{1110} & \multirow{2}{*}{14} 
& $f_0 \left( \Delta u \right) = {\Lambda_0}^{-1}$ 
& $f_1 \left( \Delta u \right) = {\Lambda_1}^{-1}$
& $f_2 \left( \Delta u \right) = {\Lambda_2}^{-1}$ 
& $f_3 \left( \Delta u \right) = {\Lambda_3}^{-1}$ 
\\ & & $\approx 1.702-i*1.702$ & $\approx -1.702-i*1.702$ & $\approx -1.702+i*1.702$ & $\approx 1.702+i*1.702$ \\
\hline \hline
\multicolumn{2}{|c||}{\multirow{2}{*}{$\Lambda$}} & $\Lambda_0 = \sqrt[4]{{\lambda}^3}$ 
& $\Lambda_1 = i*\sqrt[4]{{\lambda}^3}$ & $\Lambda_2 = -\sqrt[4]{{\lambda}^3}$ 
& $\Lambda_3 =-i* \sqrt[4]{{\lambda}^3}$ 
\\ \multicolumn{2}{|c||}{} & $\approx 0.2938+i*0.2938$ & $\approx -0.2938+i*0.2938$ & $\approx -0.2938-i*0.2938$ &  $\approx 0.2938-i*0.2938$ \\
\hline \hline
\end{tabular}
\end{center}
\caption{The nonzero transformation matrix elements of columns from 11 to 14}
\label{tab:col_N_11_to_14}
\end{table}
\endgroup

\end{enumerate}

Using Wolfram Mathematica, the authors performed a similarity transformation with the matrix constructed in this subsection. A strictly diagonal matrix was obtained, consisting of these eigenvalues.

\subsection{Generalization of determining the Jordan normal form} \label{subsec:gen_Jordan}

Of the various properties of the Jordan normal form of the elementary action $\overleftrightarrow{Z} \left( \lambda \right)$ of (\ref{eq:mat_Z}), two are distinguished for further analysis:

\begin{enumerate}

   \item There is a single dominant eigenvalue equal to 1, such that any other eigenvalue in absolute value is strictly smaller than 1. \newline
   The associated subspace is one-dimensional of vacuum eigenvector $\overleftarrow{W_{00 \ldots 0}}$ (see (\ref{eq:vacc})). \newline
   Let the dominant eigenvalue be the top left element of the Jordan normal form. So the transformation matrix $\overleftrightarrow{S_6} \left( \lambda \right)$ has the top row and the left column equal to 0, excepting the top-left element equal to 1. 
   
   \item The normal Jordan form of the elementary action $\overleftrightarrow{Z} \left( \lambda \right)$ of (\ref{eq:mat_Z}) is strictly diagonal of the eigenvalues. Let it be denoted as $\overleftrightarrow{\Lambda} \left( \lambda \right)$, thus
   
   \begin{equation} \label{eq:Jordan}
       \overleftrightarrow{\Lambda} \left( \lambda \right) = {\overleftrightarrow{S_6}}^{-1} \left( \lambda \right) * \overleftrightarrow{Z} \left( \lambda \right) * \overleftrightarrow{S_6} \left( \lambda \right).
   \end{equation}
   
   Some of the eigenvalues may coincide (see $\Lambda_0$ of Table \ref{tab:col_N_9_10} and $\Lambda_1$ of Table \ref{tab:eigenvectors_N_5}, also $\Lambda_1$ of Table \ref{tab:col_N_9_10} and $\Lambda_3$ of Table \ref{tab:eigenvectors_N_5}). But they are in different sub-subspaces, and each sub-subspace consists of different eigenvalues and is diagonalized independently.

\end{enumerate}

\subsection{The partition function and its analysis} \label{subsec:part_func}

Let the similarity transformation with the matrix $\overleftrightarrow{S_0}$ of (\ref{eq:mat_S_0}) be performed over the partition function $Z$ of (\ref{eq:Z})

\begin{equation} \label{eq:Z_trans_S0}
    Z = \left( \overrightarrow{Z_{N+1}} * \overleftrightarrow{S_0} \right) * {\left( {\overleftrightarrow{S_0}}^{-1} * \overleftrightarrow{Z_1} * \overleftrightarrow{S_0} \right)}^N * \left( {\overleftrightarrow{S_0}}^{-1} * \overleftarrow{Z_0} \right).
\end{equation}

Accounting (\ref{eq:mat_Z_1_S_0}) gives the block-diagonal form

\begin{equation} \label{eq:Z_trans_S0_block}
    Z = \left( \overrightarrow{Z_{N+1}} * \overleftrightarrow{S_0} \right) * {
    \begin{pmatrix}    
      \overleftrightarrow{B_0}_{[R]} * \overleftrightarrow{P_r} & \overleftrightarrow{0}
    \\
      \overleftrightarrow{0} & \overleftrightarrow{B_1}_{[R]} * \overleftrightarrow{P_r}
    \end{pmatrix}}^N * 
    \left( {\overleftrightarrow{S_0}}^{-1} * \overleftarrow{Z_0} \right).
\end{equation}

Let the similarity transformation with the matrix $\overleftrightarrow{S_5}$ of (\ref{eq:mat_S_5}) be performed over the partition function $Z$ of (\ref{eq:Z_trans_S0_block}). Accounting (\ref{eq:mat_Z_1_S_0_S_5}) diagonalizes blocks $\overleftrightarrow{B_0}_{[R]}$ and $\overleftrightarrow{B_1}_{[R]}$, then accounting (\ref{eq:mat_Z}) gives

\begin{equation} \label{eq:Z_actions}
    Z = \left( \overrightarrow{Z_{N+1}} * \overleftrightarrow{S_0} * \overleftrightarrow{S_5} \right) * {
    \begin{pmatrix}    
      \lambda_1 *\overleftrightarrow{Z} \left( \lambda_{21} \right) & \overleftrightarrow{0}
    \\
      \overleftrightarrow{0} & \lambda_3 *\overleftrightarrow{Z} \left( \lambda_{43} \right) 
    \end{pmatrix}}^N * 
    \left( {\overleftrightarrow{S_5}}^{-1} * {\overleftrightarrow{S_0}}^{-1} * \overleftarrow{Z_0} \right).
\end{equation}

Let $2^{R+1}\times2^{R+1}$ matrix $\overleftrightarrow{S_7}$ be constructed of two $2^R\times2^R$ matrices $\overleftrightarrow{S_6}$ of (\ref{eq:Jordan}) as follows
 
\begin{equation} \label{eq:mat_S_7}
    \overleftrightarrow{S_7} = 
    \begin{pmatrix}    
      \overleftrightarrow{S_6} \left( \lambda_{21} \right) & \overleftrightarrow{0}
    \\
      \overleftrightarrow{0} & \overleftrightarrow{S_6} \left( \lambda_{43} \right)
    \end{pmatrix}.
\end{equation}

Let the similarity transformation with the matrix $\overleftrightarrow{S_7}$ of (\ref{eq:mat_S_7}) be performed over the partition function $Z$ of (\ref{eq:Z_actions}). Accounting (\ref{eq:Jordan}) gives strictly diagonal matrices $\overleftrightarrow{\Lambda} \left( \lambda_{21} \right)$ and $\overleftrightarrow{\Lambda} \left( \lambda_{43} \right)$, then finally

\begin{equation} \label{eq:Z_diag}
    Z = \left( \overrightarrow{Z_{N+1}} * \overleftrightarrow{S_0} * \overleftrightarrow{S_5} * \overleftrightarrow{S_7} \right) * 
    \begin{pmatrix}    
      {\lambda_1}^N * {\overleftrightarrow{\Lambda}}^N \left( \lambda_{21} \right) & \overleftrightarrow{0}
    \\
      \overleftrightarrow{0} & {\lambda_3}^N *{\overleftrightarrow{\Lambda}}^N \left( \lambda_{43} \right) 
    \end{pmatrix} * 
    \left( {\overleftrightarrow{S_7}}^{-1} * {\overleftrightarrow{S_5}}^{-1} * {\overleftrightarrow{S_0}}^{-1} * \overleftarrow{Z_0} \right).
\end{equation}

Let the partition function $Z$ of (\ref{eq:Z_diag}) be analysed for a large amount of cells $N$. First, $\lambda_1$ is greater than $\lambda_3$ (see (\ref{eq:lambda})). Second, the strictly diagonal normal Jordan form $\overleftrightarrow{\Lambda}$ has a dominant eigenvalue 1 as the top-left element. Then, in a strictly diagonal matrix of (\ref{eq:Z_diag}), the top-left element ${\lambda_1}^N$ is much larger than the rest. Third, the transformation matrix $\overleftrightarrow{S_6} \left( \lambda \right)$ has the top row and the left column equal to 0, excepting the top-left element equal to 1, and this is also true for $\overleftrightarrow{S_7}$ due to (\ref{eq:mat_S_7}). Then the partition function $Z$ is close to

\begin{equation} \label{eq:Z_infin}
    Z_{\infty} = {\lambda_1}^N * \left( \overrightarrow{Z_{N+1}} * \overleftrightarrow{S_0} * \overleftrightarrow{S_5} \right)_0 * 
    \left( {\overleftrightarrow{S_5}}^{-1} * {\overleftrightarrow{S_0}}^{-1} * \overleftarrow{Z_0} \right)_0,
\end{equation}

where $\displaystyle \left( \overrightarrow{Z_{N+1}} * \overleftrightarrow{S_0} * \overleftrightarrow{S_5} \right)_0$ is the left element numbered 0 of the row vector $\displaystyle \left( \overrightarrow{Z_{N+1}} * \overleftrightarrow{S_0} * \overleftrightarrow{S_5} \right)$, and $\displaystyle \left( {\overleftrightarrow{S_5}}^{-1} * {\overleftrightarrow{S_0}}^{-1} * \overleftarrow{Z_0} \right)_0$ is the top element numbered 0 of the column vector $\displaystyle \left( {\overleftrightarrow{S_5}}^{-1} * {\overleftrightarrow{S_0}}^{-1} * \overleftarrow{Z_0} \right)$.

Note that (\ref{eq:Z_infin}) is the same as (38) of \cite{sakhno2021free}.

From the partition function one gets the free energy $A$ and the specific free energy per spin $a$ (see (39) of \cite{sakhno2021free})

\begin{equation} \label{eq:free_en}
    A=-k_B*T*ln(Z), \quad a=\frac{A}{2*N+R+1},
\end{equation}

where $2*N+R+1$ is the amount of spins, since each of the $N$ internal cells has 2 spins and the finish cell has $R+1$ spins with spin first sub-number equal to the cell number.

Taking into account (\ref{eq:Z_infin}), with a large amount of cells $N$, the free energy $A$ and the specific free energy per spin $a$ of (\ref{eq:free_en}) are close to

\begin{equation} \label{eq:free_en_inf}
    A_{\infty}=-k_B * T * N * \ln(\lambda_1), \quad 
    a_{\infty}=-k_B * T * \cfrac{\ln(\lambda_1)}{2}.
\end{equation}

It is important that the properties of (\ref{eq:free_en_inf}) do not depend on the number of rows $R$ and the boundary conditions.

\section{Conclusion}

\begin{enumerate}
   \item The simplest quantum field model is built in Section \ref{sec:model} for the block-diagonalizable two-dimensional generalized Ising systems, introduced in \cite{sakhno2020matrix} and \cite{sakhno2021free}. In this model, the elementary action operator of (\ref{eq:mat_Z}) is introduced, which is analyzed further in the paper.
   \item The eigenvalues of the elementary action were analyzed in Section \ref{sec:eigenvalues}. They may be calculated by (\ref{eq:Lam_roots}).
   \item The eigenvectors of the elementary action were analyzed in Section \ref{sec:eigenvectors}. Their coordinates may be calculated by (\ref{eq:fs_gen_gen}) and (\ref{eq:wave_func}).
   \item It's funny that the simplest quantum field model yields the simplest models of: the special relativity, the general relativity, the uncertainty principle,  the wave-particle duality (see Subsection \ref{subsec:fun}).
   \item The Jordan normal form of the elementary action was analyzed in Section \ref{sec:Jordan}. It has a single dominant eigenvalue and is strictly diagonal of the eigenvalues, although some of the eigenvalues may coincide.
   \item The partition function was analyzed in Subsection \ref{subsec:part_func}. With a large amount of cells the specific free energy per spin does not depend on the number of rows and the boundary conditions (see (\ref{eq:free_en_inf})).
\end{enumerate}

\bibliographystyle{unsrtnat}
\bibliography{main}

\end{document}